\documentclass[12pt,preprint]{aastex}

\newtheorem{theorem}{Theorem}

\newenvironment{proof}{~ \\[0.1in] {\bf Proof.} }{\hfill $\Box$ \bigskip \\[0.1in]}

\usepackage[intlimits]{amsmath}
\usepackage{natbib}
\usepackage{graphicx}
\usepackage{amssymb}
\newcommand{\refeqn}[1]{Eq. (\ref{#1})}

\newcommand{\mysubsection}[1]{{\bf {#1}}}

\newcommand{\xt}{\tilde{x}}
\newcommand{\yt}{\tilde{y}}
\newcommand{\rt}{\tilde{r}}
\newcommand{\at}{\tilde{a}}

\newcommand{\htild}{\tilde{h}}

\newcommand{\Rt}{\tilde{R}}
\newcommand{\Apupmap}{A_{\mathrm{pupmap}}}
\newcommand{\Ain}{A_{\mathrm{in}}}
\newcommand{\Aout}{A_{\mathrm{out}}}
\newcommand{\Ein}{E_{\mathrm{in}}}
\newcommand{\Ei}{E_{\mathrm{i}}}

\newcommand{\Eout}{E_{\mathrm{out}}}
\newcommand{\Eo}{E_{\mathrm{o}}}

\newcommand{\thetat}{\tilde{\theta}}

\shorttitle{Sensitivity Analysis for Pupil Mapping}
\shortauthors{Belikov et. al.}

\begin{document}


\title{Diffraction-Based Sensitivity Analysis of Apodized Pupil Mapping Systems}

\author{Ruslan Belikov}
\affil{Mechanical and Aerospace Engineering, Princeton University}
\email{rbelikov@princeton.edu}

\author{N. Jeremy Kasdin}
\affil{Mechanical and Aerospace Engineering, Princeton University}
\email{jkasdin@princeton.edu}

\author{Robert J. Vanderbei}
\affil{Operations Research and Financial Engineering, Princeton University}
\email{rvdb@princeton.edu}

%

\begin{abstract}
Pupil mapping is a promising and unconventional new method for high
contrast imaging  being considered for terrestrial exoplanet
searches. It employs two (or more) specially designed aspheric
mirrors to create a high-contrast amplitude profile across the telescope
pupil that does not appreciably attenuate amplitude.  As such, it
reaps significant benefits in light collecting efficiency and inner
working angle, both critical parameters for terrestrial planet
detection.  While much has been published on various aspects of
pupil mapping systems, the problem of sensitivity to wavefront
aberrations remains an open question. In this paper, we present an
efficient method for computing the sensitivity of a pupil mapped
system to Zernike aberrations. We then use this method to study the
sensitivity of a particular pupil mapping system and compare it to
the concentric-ring shaped pupil coronagraph. In particular, we quantify
how contrast and inner working angle degrade with increasing Zernike
order and rms amplitude.  These results have obvious ramifications for the
stability requirements and overall design of a planet-finding observatory.


\end{abstract}


\section{Introduction} \label{sec:Introduction}

The impressive discoveries of large extrasolar planets over the past
decade have inspired widespread interest in finding and directly
imaging Earth-like planets in the habitable zones of nearby stars.
In fact, NASA has plans to launch two space telescopes to accomplish
this,  the {\em Terrestrial Planet Finder Coronagraph (TPF-C)} and
the {\em Terrestrial Planet Finder Interferometer (TPF-I)}, while
the European Space Agency is planning a similar multi-satellite
mission called Darwin. These missions are currently in the concept
study phase.  In addition, numerous ground-based searches are
proceeding using both coronagraphic and interferometric approaches.

Direct imaging of Earth-like extrasolar planets in the habitable zones of
Sun-like stars poses an extremely challenging problem in high-contrast
imaging.  Such a star will shine $10^{10}$ times more brightly than the
planet.  And, if we assume that the star-planet system is $10$ parsecs from us,
the maximum separation between the star and the planet will be roughly $0.1$
arcseconds.

\mysubsection{Design Concepts for TPF-C.} For TPF-C, for example,
the current baseline design involves a traditional {\em Lyot
coronagraph} consisting of a modern 8th-order occulting mask (see,
e.g., \citet{KCG04}) attached to the back end of a Ritchey-Chretien
telescope having an $8$m by $3.5$m elliptical primary mirror.
Alternative innovative back-end designs still being considered
include {\em shaped pupils} (see, e.g., \citet{KVSL02} and
\citet{VKS04}), a {\em visible nuller} (see, e.g., \citet{SLSWL04})
and {\em pupil mapping} (see, e.g., \citet{Guy03} where this
technique is called {\em phase-induced amplitude apodization} or
{\em PIAA}). By pupil mapping we mean a system of two lenses, or
mirrors, that takes a flat input field at the entrance pupil and
produce an output field that is amplitude modified but still flat in
phase (at least for on-axis sources).

\mysubsection{The Pupil Mapping Concept.}
The pupil mapping concept has received considerable attention recently because
of its high throughput and small effective inner working angle
(IWA). These benefits could potentially permit more observations
over the mission lifetime, or conversely, a smaller and cheaper
overall telescope. As a result, there have been numerous studies
over the past few years to examine the performance of pupil mapping
systems. In particular, \citet{Guy03,TV03,VT04,GPGMRW05} derived expressions
for the optical surfaces using ray optics. However, these
analyses made no attempt to provide a complete diffraction through a
pupil mapping system. More recently, \citet{Van05} provided a
detailed diffraction analysis. Unfortunately, this analysis showed
that a pupil mapping system, in its simplest and most elegant form,
cannot achieve the required
$10^{-10}$ contrast;  the diffraction effects from the pupil mapping
systems themselves are so detrimental that contrast is limited to $10^{-5}$.
In \cite{GPGMRW05} and
\citet{PGRMWBG06}, a {\em hybrid pupil mapping} system was proposed that
combines the pupil mapping mirrors with a modest apodization of oversized
entrance and exit pupils. This combination does indeed achieve the
needed high-contrast point spread function (PSF).
In this paper, we call such systems {\em apodized pupil mapping} systems.

A second problem that must be addressed is the fact that a simple
two-mirror (or two-lens) pupil mapping system introduces
non-constant angular magnification for off-axis sources (such as a
planet).  In fact, the off-axis magnification for light passing
through a small area of the exit pupil is directly proportional to
the amplitude amplification in that small area. For systems in which
the exit amplitude amplification is constant, the magnification is
also constant. But, for high-contrast imaging, we are interested in
amplitude profiles that are far from constant.  Hence, off-axis
sources do not form images in a formal sense (the "images" are very
distorted.) \citet{Guy03} proposed an elegant solution to this
problem. He suggested using this system merely as a mechanism for
concentrating (on-axis) starlight in an image plane. He then
proposed that an occulter be placed in the image plane to remove the
starlight. All other light, such as the distorted off-axis planet
light, would be allowed to pass through the image plane. On the back
side would be a second, identical pupil mapping system  (with the
apodizers removed), that would ``umap'' the off-axis beam and thus
remove the distortions introduced by the first system (except for
some beam walk---see \citet{VT04}).

\mysubsection{Sensitivity Analysis.} What remains to be answered is
how apodized pupil mapping behaves in the presence of optical
aberrations.  It is essential that contrast be maintained during an
observation, which might take hours during which the wavefront will
undoubtedly suffer aberration due to the small dynamic perturbations
of the primary mirror.  An understanding of this sensitivity is
critical to the design of TPF-C or any other observatory. In
\citet{Green04}, a detailed sensitivity analysis is given for shaped
pupils and various Lyot coronagraphs (including the $8$th-order
image plane mask introduced in \citet{KCG04}). Both of these design
approaches achieve the needed sensitivity for a realizable mission.
So far, however, no comparable study has been done for apodized
pupil mapping. One obstacle to such a study is the considerable
computing power required to do a full 2-D diffraction simulation.

\mysubsection{Aberrations Given by Zernike Polynomials.} In this
paper, we present an efficient method for computing the effects of
wavefront aberrations on apodized pupil mapping. We begin with a
brief review  of the design of apodized pupil mapping systems in
Section \ref{sec:Review of Pupil Mapping and Apodization}. We then
present in Section \ref{sec:Diffraction Analysis} a semi-analytical
approach to computing the PSF of systems such as pupil-mapping and
concentric rings in the presence of aberrations represented by
Zernike polynomials. For such aberrations, it is possible to
integrate analytically the integral over azimuthal angle, thereby
reducing the computational problem from a double integral to a
single one, eliminating the need for massive computing power.

In Section \ref{sec:Simulations}, we present the sensitivity results
for an apodized pupil mapping system and a concentric ring shaped
pupil coronagraph, and compare the results.

\section{Review of Pupil Mapping and Apodization}
\label{sec:Review of Pupil Mapping and Apodization}

In this section, we review the apodized pupil mapping approach and
introduce the specific system that we study in subsequent sections.
It should be noted that this apodized pupil mapping design may not
be the best possible. Rather, it is merely an example of such a
system that achieves high contrast. Other examples can be found in
the recent paper by \citet{PGRMWBG06}. Our aim in this paper is not
to identify the best such system. Instead, our aim is to develop
tools for carrying a full diffraction analysis of any apodized pupil
mapping system.

\subsection{Pupil Mapping via Ray Optics}
\label{subsec:Pupil Mapping via Ray Optics}
\begin{figure}
\begin{center}
\mbox{\includegraphics[width=2.5in]{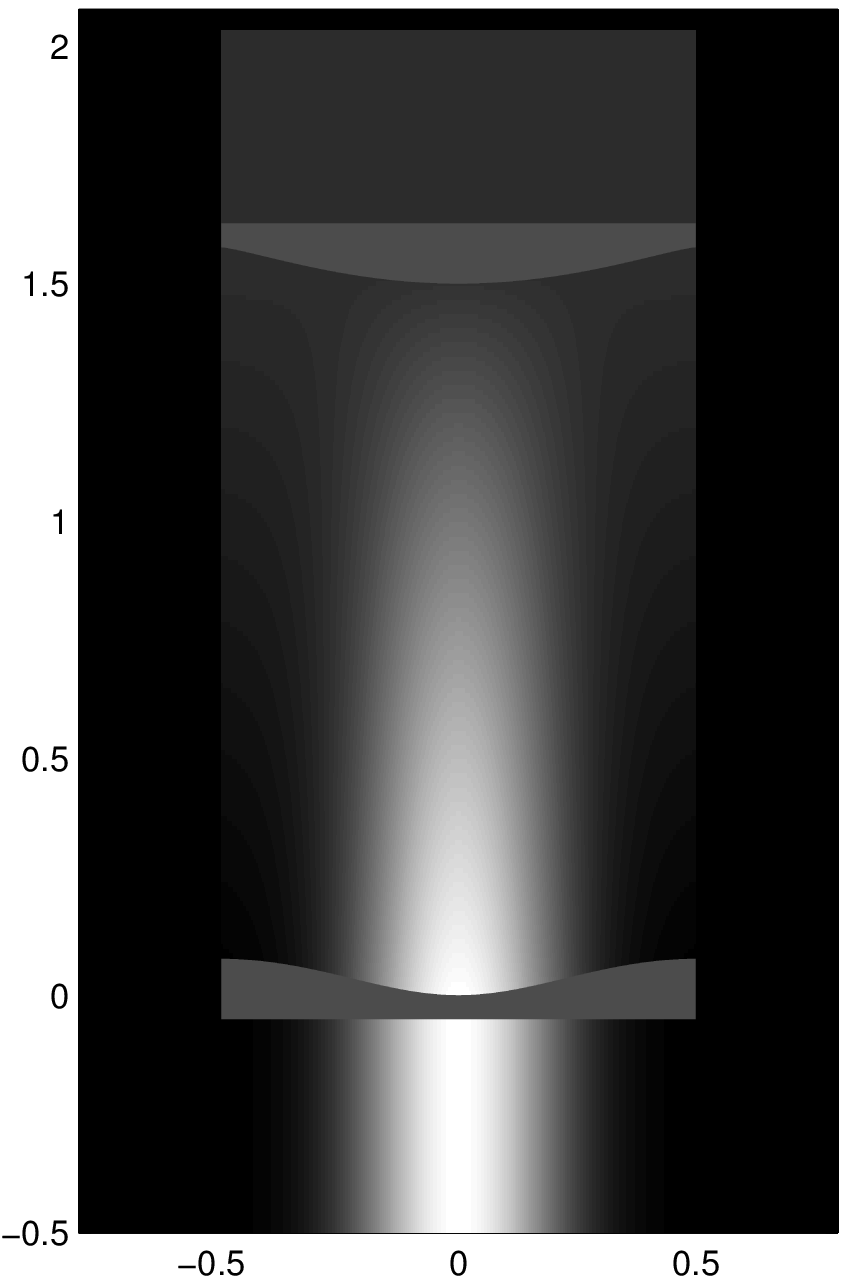}}
\mbox{\includegraphics[width=3.2in]{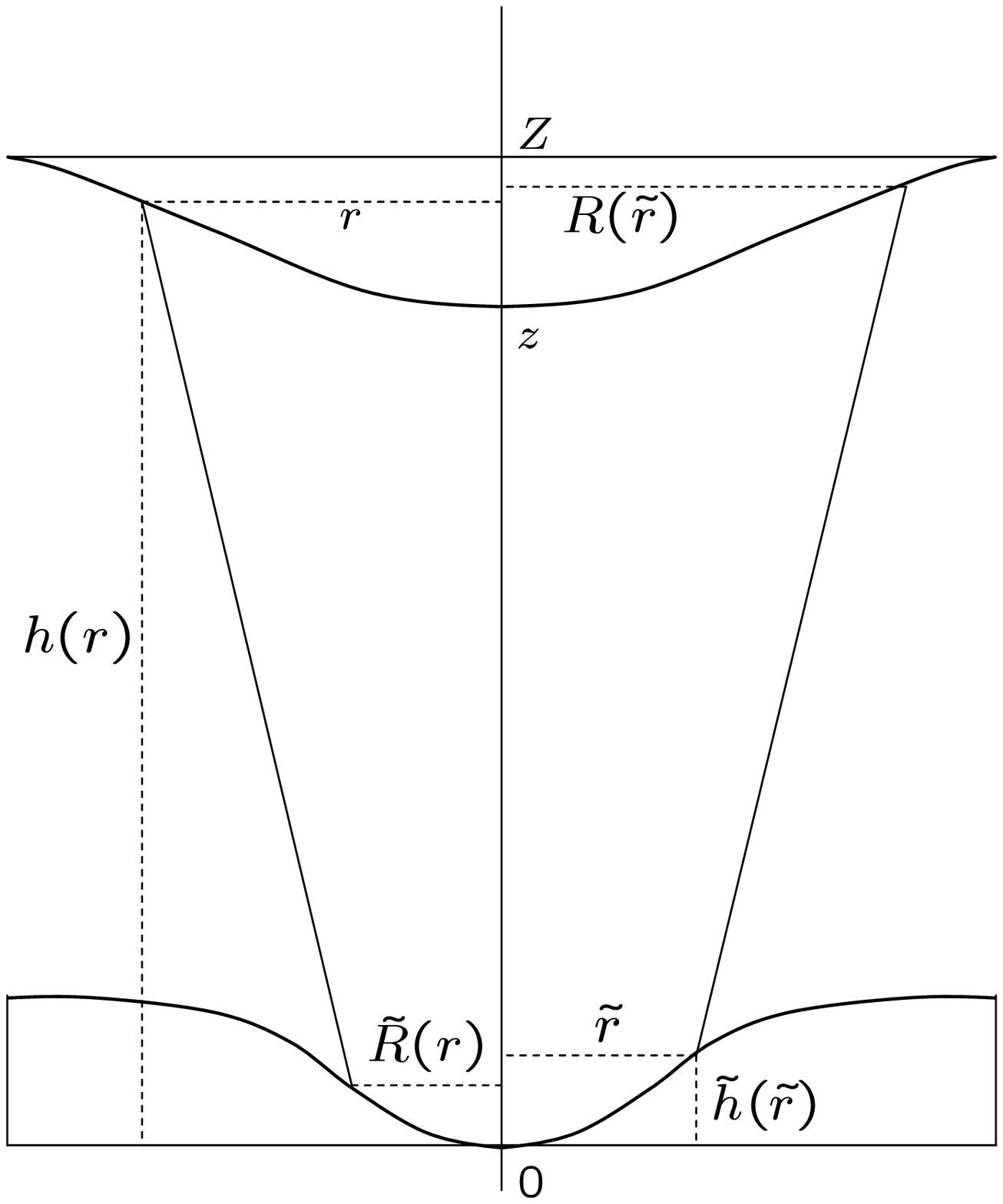}}
\end{center}
\caption{Pupil mapping via a pair of properly figured lenses. Light
travels from top to bottom.} \label{fig:1}
\end{figure}

We begin by summarizing the ray-optics description of pure pupil mapping.
An on-axis ray entering the first pupil at radius $r$ from the
center is to be mapped to radius $\rt = \Rt(r)$ at the exit pupil (see Figure
\ref{fig:1}).
Optical elements at the two pupils ensure that the exit ray is
parallel to the entering ray. The function $\Rt(r)$ is assumed to be
positive and increasing or, sometimes, negative and decreasing.  In
either case, the function has an inverse that allows us to recapture
$r$ as a function of $\rt$: $r = R(\rt)$. The purpose of pupil
mapping is to create nontrivial amplitude profiles. An amplitude
profile function $A(\rt)$ specifies the ratio between the output
amplitude at $\rt$ to the input amplitude at $r$ (in a pure
pupil-mapping system the input amplitude is constant).
\citet{VT04} showed that for any desired amplitude profile $A(\rt)$ there is a
pupil mapping function $R(\rt)$ that achieves it (in a ray-optics
sense).  Specifically, the pupil mapping is given by
\begin{equation}\label{1}
    R(\rt) = \pm \sqrt{\int_0^{\rt} 2 A^2(s) s ds} .
\end{equation}
Furthermore, if we consider the case of a pair of lenses that are planar on
their outward-facing surfaces, 
then the inward-facing surface profiles, $h(r)$ and $\htild(\rt)$, that
are required to obtain the desired pupil mapping are given by the solutions to
the following ordinary differential equations:
\begin{equation}\label{2}
    \frac{\partial h}{\partial r}(r)
    = \frac{r-\Rt(r)}{ |n-1| \sqrt{z^2
        + \frac{n+1}{n-1}(r-\Rt(r))^2} },
    \qquad
    h(0) = z,
\end{equation}
and
\begin{equation}\label{3}
    \frac{\partial \htild}{\partial \rt}(\rt)
    = \frac{R(\rt)-\rt}{ |n-1| \sqrt{\phantom{\tilde{|}}z^2
        + \frac{n+1}{n-1}(R(\rt)-\rt)^2} },
    \qquad
    \htild(0) = 0.
\end{equation}
Here, $n \ne 1$ is the refractive index and $z$ is the distance separating
the centers ($r=0$, $\rt = 0$) of the two lenses.

Let $S(r,\rt)$ denote the distance between a point on the
first lens surface $r$ units from the center and the corresponding point on
the second lens surface $\rt$ units from its center.  Up to an additive
constant, the optical path length of a ray that exits at radius
$\rt$ after entering at radius $r = R(\rt)$ is given by
\begin{equation}\label{4}
    Q_0(\rt) = S(R(\rt),\rt) + |n|(\htild(\rt)-h(R(\rt))).
\end{equation}
\citet{VT04} showed that, for an on-axis source,
$Q_0(\rt)$ is constant and equal to $-(n-1)|z|$.%
\footnote{ For a pair of mirrors, put $n=-1$.  In that case, $z<0$ as the
first mirror is ``below'' the second.  }

\subsection{High-Contrast Amplitude Profiles}
\label{subsec:High-Contrast Amplitude Profiles}

\begin{figure}
\begin{center}
\mbox{\includegraphics[width=3.0in]{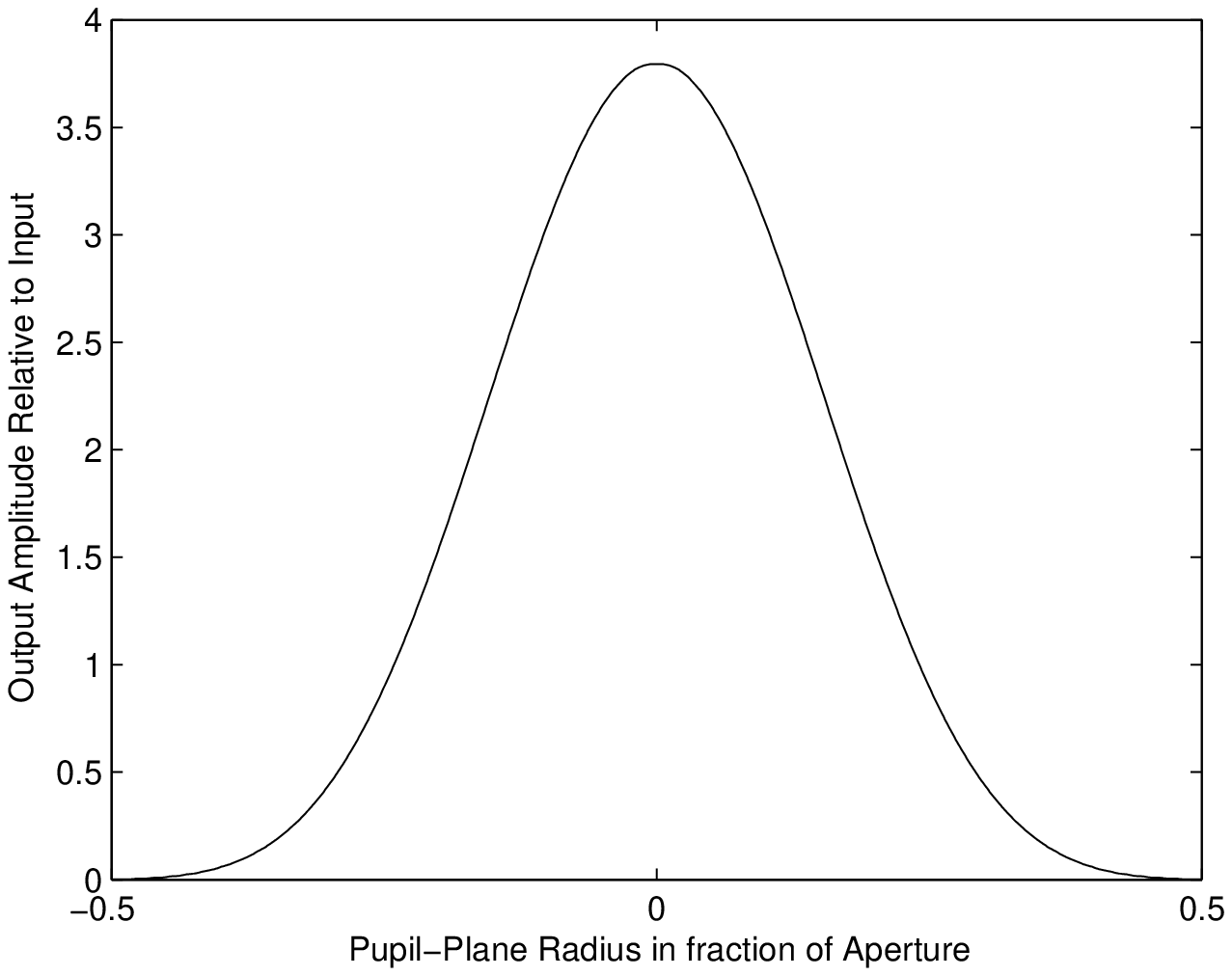}}
\mbox{\includegraphics[width=3.0in]{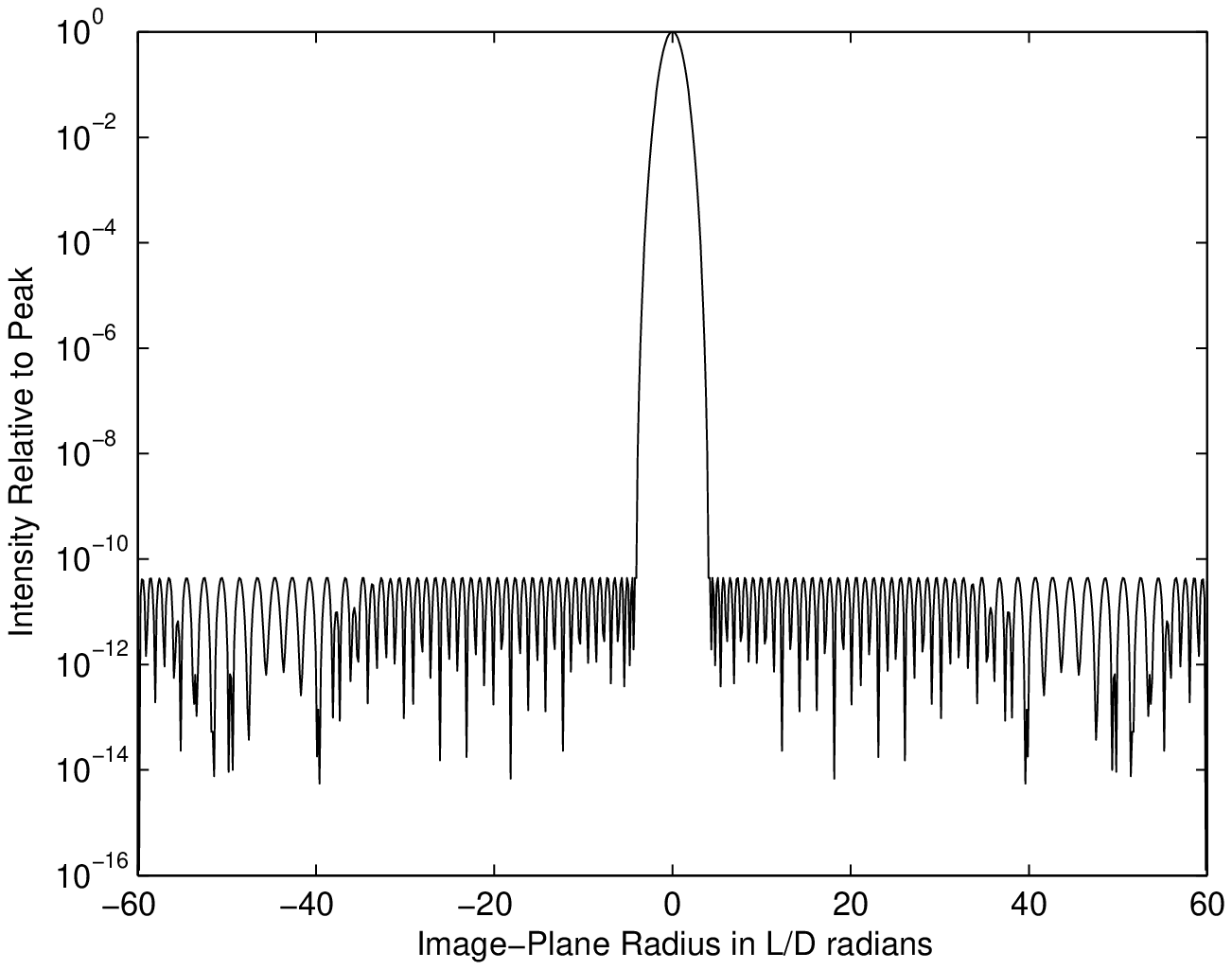}}
\end{center}
\caption{{\em Left.}
An amplitude profile providing contrast of $10^{-10}$ at tight inner working
angles.
{\em Right.} The corresponding on-axis point spread function.}
\label{fig:2}
\end{figure}

If we assume that a collimated beam with amplitude profile $A(\rt)$
such as one obtains as the output of a pupil mapping system
is passed into an ideal imaging system with focal length $f$, the
electric field $E(\rho)$ at the image plane is given by
the Fourier transform of $A(\rt)$:
\begin{equation}\label{5}
    E(\xi,\eta)
    =
    \frac{E_0}{\lambda i f}
    e^{\pi i \frac{\xi^2 + \eta^2}{\lambda f}}
    \int_{-\infty}^{\infty}\int_{-\infty}^{\infty}
    e^{-2 \pi i \frac{\xt\xi + \yt\eta}{\lambda f}}
          A(\sqrt{\xt^2+\yt^2}) d\yt d\xt .
\end{equation}
Here, $E_0$ is the input amplitude which, unless otherwise noted,
we take to be unity.
Since the optics are azimuthally symmetric, it is convenient to use polar
coordinates.  The amplitude profile $A$ is a function of
$\rt = \sqrt{\xt^2+\yt^2}$ and the image-plane electric field depends only on
image-plane radius $\rho = \sqrt{\xi^2 + \eta^2}$:
\begin{eqnarray}
    E(\rho)
    & = &
    \frac{1}{\lambda i f}
    e^{\pi i \frac{\xi^2 + \eta^2}{\lambda f}}
    \int_0^{\infty}\int_0^{2 \pi}
        e^{-2 \pi i \frac{\rt \rho}{\lambda f} \cos(\theta - \phi)}
                A(\rt) \rt d\theta d\rt \label{6} \\
    & = &
    \frac{2 \pi}{\lambda i f}
    e^{\pi i \frac{\xi^2 + \eta^2}{\lambda f}}
    \int_0^{\infty} J_0\left(-2 \pi \frac{\rt \rho}{\lambda f}\right)
               A(\rt) \rt d\rt . \label{7}
\end{eqnarray}
The point-spread function (PSF) is the square of the electric field:
\begin{equation}\label{8}
    \mbox{Psf}(\rho) = |E(\rho)|^2 .
\end{equation}
For the purpose of terrestrial planet finding, it is important to construct an
amplitude profile for which the PSF at small nonzero angles is ten orders of
magnitude reduced from its value at zero.
A paper by \citet{VSK03} explains how these functions are computed as
solutions to certain optimization problems.
The high-contrast amplitude profile used in the rest of this paper is shown in
Figure \ref{fig:2}.

\subsection{Apodized Pupil Mapping Systems} \label{sec:hybrid}
\label{subsec:Apodized Pupil Mapping Systems}

\citet{Van05} showed that pure pupil mapping systems designed for
contrast of $10^{-10}$ actually achieve much less than this due to
harmful diffraction effects that are not captured by the simple ray
tracing analysis outlined in the previous section.  For most systems
of practical real-world interest (i.e., systems with apertures of a
few inches and designed for visible light), contrast is limited to
about $10^{-5}$.  \citet{Van05} considered certain hybrid designs
that improve on this level of performance but none of the
hybrid designs presented there completely overcame this
diffraction-induced contrast degradation.

In this section, we describe an apodized pupil mapping system that
is somewhat more complicated than the designs presented in
\citet{Van05}. This hybrid design, based on ideas proposed by
Olivier Guyon and Eugene Pluzhnik (see \citet{PGRMWBG06}), involves
three additional components.  They are
\begin{enumerate}
    \item a preapodizer $A_0$
    to soften the edge of the first lens/mirror so as to
    minimize diffraction effects caused by hard edges,
    \item a postapodizer to smooth out low spatial frequency ripples produced
    by diffraction effects induced by the pupil mapping system itself, and
    \item a backend phase shifter to smooth out low spatial frequency ripples
    in phase.
\end{enumerate}
Note that the backend phase shifter can be built into the second
lens/mirror. There are several choices for the preapodizer. For this
paper, we choose Eqs. (3) and (4) in \citet{PGRMWBG06} for our
pre-apodizer:
\[
    A_0(r) = \frac{A(r)(1 + \beta)}{A(r) + \beta A_{\mathrm{max}}},
\]
where $A_{\mathrm{max}}$ denotes the maximum value of $A(r)$ and $\beta$ is a
scalar parameter, which we take to be $0.1$.  It is easy to see that
\begin{itemize}
    \item $A(r)/A_{\mathrm{max}} \le A_0(r) \le 1$ for all $r$,
    \item $A_0(r)$ approaches $1$ as $A(r)$ approaches $A_{\mathrm{max}}$, and
    \item $A_0(r)$ approaches $0$ as $A(r)$ approaches $0$.
\end{itemize}

Incorporating a post-apodizer introduces a degree of freedom
that is lacking in a pure pupil mapping system.  Namely, it is possible to
design the pupil mapping system based on an arbitrary amplitude profile and
then convert this profile to a high-contrast profile via an
appropriate choice of backend
apodizer.  We have found that a simple Gaussian amplitude profile that
approximately matches a high-contrast profile works very well.  Specifically,
we used
\[
    \Apupmap(\rt) = 3.35 e^{-22(\rt/\at)^2},
\]
where $\at$ denotes the radius of the second lens/mirror.

The backend apodization is computed by taking the actual output
amplitude profile as computed by a careful diffraction analysis,
smoothing it by convolution with a Gaussian distribution, and then
apodizing according to the ratio of the desired high-contrast
amplitude profile $A(\rt)$ divided by
the smoothed output profile.  Of course, since a true apodization can never
intensify a beam, this ratio must be further scaled down so that it is nowhere
greater than unity.  The Gaussian convolution kernel we used has mean zero and
standard deviation $\at/\sqrt{100,000}$.

The backend phase modification is computed by a similar smoothing operation
applied to the output phase profile.  Of course, the smoothed output phase
profile (measured in radians) must be converted to a surface profile
(having units of length).  This conversion requires us
to assume a certain specific wavelength.  As a consequence, the resulting
design is correct only at one wavelength.  The ability of the system
to achieve high contrast
degrades as one moves away from the design wavelength.

\subsection{Star Occulter and Reversed System}
\label{subsec:Star Occulter and Reversed System}

It is important to note that the PSFs in Figure \ref{fig:2} correspond to a
bright on-axis source (i.e., a star).  Off-axis sources, such as faint
planets, undergo two effects in a pupil mapping system that differ from the
response of a conventional imaging system: an effective magnification and a
distortion. These are explained in detail in \citet{VT04} and \citet{TV03}.
The magnification, in particular, is due to an overall narrowing of the exit
pupil as compared to the entrance pupil.  It is this magnification that
provides pupil mapped systems their smaller effective inner working angle.
The techniques in Section \ref{sec:Diffraction Analysis}
will allow us to compute the exact
off-axis diffraction pattern of an apodized pupil mapped coronagraph and thus
to see these effects.

\begin{figure}
\begin{center}
\mbox{\includegraphics[width=6.5in]{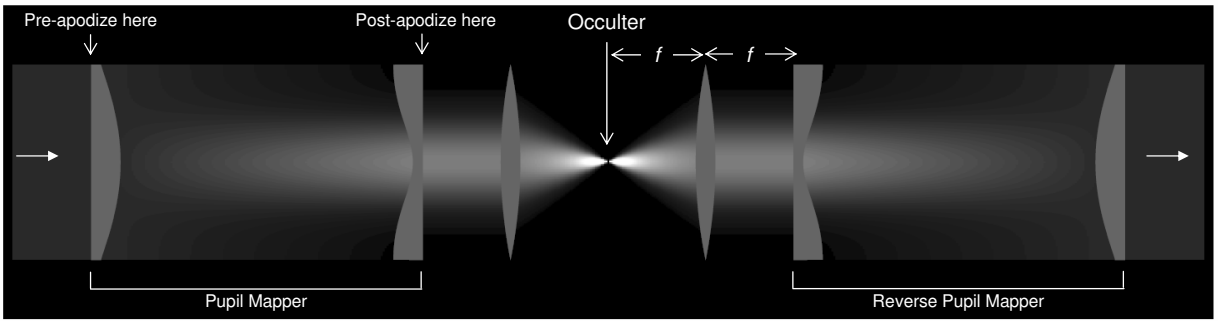}}
\end{center}
\caption{The full pupil mapping system includes:
a pair of lenses to shape the
amplitude into a prolate-spheroidal-like amplitude profile,
a focusing lens that concentrates the on-axis starlight into a small central
lobe where a (not-depicted) occulter can block this light,
followed by a recollimating lens and finally a reverse pupil mapping system
that reforms the pupil with the starlight removed but any planet light (if
present) intact.
This final pupil is then fed a final focusing element (not shown) to form an
image of off-axis sources.
}
\label{fig:3}
\end{figure}

While the effective magnification of a pupil mapping system results in an
inner working angle advantage of about a factor of two,  it does not produce
high-quaity  diffraction limited images of off-axis sources because of the
distortion inherent in the system. \citet{Guy03} proposed the following
solution to this problem. He suggested using this system merely as a mechanism
for concentrating (on-axis) starlight in an image plane. He then proposed that
an occulter be placed in the image plane to remove the starlight. All other
light, such as the distorted off-axis planet light, would be allowed to pass
through the image plane. On the back side would be a second, identical pupil
mapping system  (with the apodizers removed), that would ``umap'' the off-axis
beam and thus remove the distortions introduced by the first system (except
for some beam walk---see \citet{VT04}).  A schematic of the
full system (without the occulter) is shown in Figure \ref{fig:3}.
Note that we have spaced the lenses one focal length from the flat sides of
the two lenses.  As noted in \cite{VT04}, such a spacing guarantees that these
two flat surfaces form a conjugate pair of pupils.

\section{Diffraction Analysis}
\label{sec:Diffraction Analysis}

In \cite{Van05}, it was shown that a simple Fresnel analysis is inadequate for
validating the high-contrast imaging capabilities we seek.  Hence, a more
accurate approximation was presented.  In this section,
we give a similar but slightly different approximation that is just as
effective for studying pupil mapping but is better suited to the
full system we wish to analyze.

\subsection{Propagation of General Wavefronts}
\label{subsec:Propagation of General Wavefronts}

The goal of this section is to derive an integral
that describes how to propagate a scalar electric field from one
plane perpendicular to the direction of propagation to another
parallel plane positioned downstream of the first.  We assume that
the electric field passes through a lens at the first plane,
then propagates through free space until reaching a second lens at the second
plane through which it passes.  In order to cover the apodized pupil mapping
case discussed in the previous section, we allow both the entrance and exit
fields to be apodized.

Suppose that the input field at the first plane is $\Ein(x,y)$.
Then the electric field at a particular point
on the second plane can be well-approximated by superimposing the
phase-shifted waves from each point across the entrance pupil (this
is the well-known Huygens-Fresnel principle---see, e.g., Section 8.2
in \cite{BW99}).  If we assume that the two lenses are given by radial
``height'' functions $h(r)$ and $\htild(\rt)$, then we can write the exit
field as
\begin{equation}\label{9}
  \Eout(\xt, \yt)
  =
  \Aout(\rt)
  \int_{-\infty}^{\infty}\int_{-\infty}^{\infty}
       \frac{1}{\lambda i Q(\xt,\yt,x,y)}
       e^{ 2 \pi i Q(\xt,\yt,x,y) /\lambda }
       \Ain(r) \Ein(x,y)
       dydx,
\end{equation}
where
\begin{equation}\label{20}
       Q(\xt,\yt,x,y) =
       \sqrt{(x-\xt)^2+(y-\yt)^2+(h(r)-\htild(\rt))^2}
       + n(Z - h(r) + \htild(\rt))
\end{equation}
is the optical path length, $Z$ is the distance between the planar
lens surfaces, $\Ain(r)$
denotes the input amplitude apodization at radius $r$, $\Aout(\rt)$
denotes the output amplitude apodization at radius $\rt$, and where,
of course, we have used $r$ and $\rt$ as shorthands for the radii in
the entrance and exit planes, respectively.

As before, it is convenient to work in polar coordinates:
\begin{equation}\label{10}
  \Eout(\rt,\thetat)
  =
  \Aout(\rt)
  \int_0^{\infty}\int_0^{2 \pi}
    \frac{1}{\lambda i Q(\rt,r,\theta-\thetat)}
        e^{ 2 \pi i Q(\rt,r,\theta-\thetat)/\lambda)}
        \Ain(r) \Ein(r,\theta)
    r d\theta dr,
\end{equation}
where
\begin{equation}\label{21}
    Q(\rt,r,\theta) =
    \sqrt{r^2-2r\rt\cos\theta+\rt^2+(h(r)-\htild(\rt))^2}
        + n(Z - h(r) + \htild(\rt)) .
\end{equation}
For numerical tractability, it is essential to make approximations
so that the integral over $\theta$ can be carried out analytically, thereby
reducing the double integral to a single one.
To this end, we need to make an appropriate approximation to the
square root term:
\begin{equation}\label{13}
     S = \sqrt{r^2-2r\rt\cos\theta+\rt^2+(h(r)-\htild(\rt))^2} .
\end{equation}

A simple crude approximation is adequate for the
$1/Q(\rt,r,\theta-\thetat)$ amplitude-reduction
factor in \refeqn{10}.  We approximate this factor
by the constant $1/Z$.

The $Q(\rt,r,\theta-\thetat)$ appearing in the
exponential must, on the other hand, be treated with care.
The classical Fresnel approximation is to replace $S$ by the first two terms
in a Taylor series expansion of the square root function about
$(h(r)-\htild(\rt))^2$.  As we already mentioned, this approximation is too
crude.  It is critically important that the integrand be exactly correct when
the pair $(r,\rt)$ correspond to rays of ray optics.
Here is a method that does this.
First, we add and subtract $S(\rt,r,0)$ from $Q(\rt,r,\theta)$ in
\refeqn{10} to get
\begin{eqnarray}
    Q(\rt,r,\theta-\thetat)
    & = &
    S(\rt,r,\theta-\thetat)-S(\rt,r,0)+S(\rt,r,0)
    + |n|\left(\htild(\rt)-h(r)\right)
    \nonumber \\
    & = &
    \frac{S(\rt,r,\theta-\thetat)^2-S(\rt,r,0)^2}{
          S(\rt,r,\theta-\thetat)+S(\rt,r,0)}
    +S(\rt,r,0) + |n|\left(\htild(\rt)-h(r)\right)
    \nonumber \\
    & = &
    \frac{r \rt - r \rt \cos(\theta-\thetat)}{
          (S(\rt,r,\theta-\thetat)+S(\rt,r,0))/2}
    +S(\rt,r,0) + |n|\left(\htild(\rt)-h(r)\right) .
    \label{141}
\end{eqnarray}
So far, these calculations are exact.
The only approximation we now make is to replace
$S(\rt,r,\theta-\thetat)$ in the denominator of \refeqn{141} with
$S(\rt,r,0)$ so that the denominator becomes just $S(\rt,r,0)$.
Putting this all together, we get a new
approximation, which we refer to as the {\em S-Huygens} approximation:
\begin{eqnarray}
  \Eout(\rt,\thetat)
  & \approx &
    \frac{\Aout(\rt)}{\lambda i Z}
    \int_0^{\infty} K(r,\rt)
    \int_0^{2 \pi}
        e^{2 \pi i \left( -\frac{\rt r \cos(\theta - \thetat)}{S(\rt,r,0)}
                      \right) /\lambda} \Ein(r,\theta) d\theta
              \; \Ain(r) r dr , \label{160}
\end{eqnarray}
where
\begin{equation}
    K(r,\rt)
    =
    e^{ 2 \pi i \left(
      \frac{ r \rt }{S(\rt,r,0)} + S(\rt,r,0) + |n|(\htild(\rt) - h(r))
    \right) /\lambda}
\end{equation}
(note that we have dropped an $exp(2 \pi i n Z / \lambda)$ factor since this
factor is just a constant unit complex number which would disappear anyway at
the end when we compute intensities).

The only reason for making approximations to the Huygens-Fresnel integral
\eqref{9} is to simplify the dependence on $\theta$ so that the integral over
this variable can be carried out analytically.  For example, if we now assume
that the input field $\Ein(r,\theta)$ does not depend on $\theta$,
then the inner integral can be evaluated explicitly and we get
\begin{eqnarray}
  \Eout(\rt,\thetat)
  & \approx &
    \frac{2 \pi \Aout(\rt)}{\lambda i Z}
    \int_0^{\infty} K(r,\rt)
    J_0\left( \frac{2 \pi \rt r}{\lambda S(\rt,r,0)} \right)
              \Ein(r) \; \Ain(r) r dr , \label{161}
\end{eqnarray}
Removing the dependency on $\theta$ greatly simplifies computations
because we only need to compute a 1D integral instead of 2D. In the
next subsection we will show how to achieve similar reductions in
cases where the dependence of $\Ein$ on $\theta$ takes a simple
form.

\begin{figure}
\begin{center}
\mbox{\includegraphics[width=6.5in]{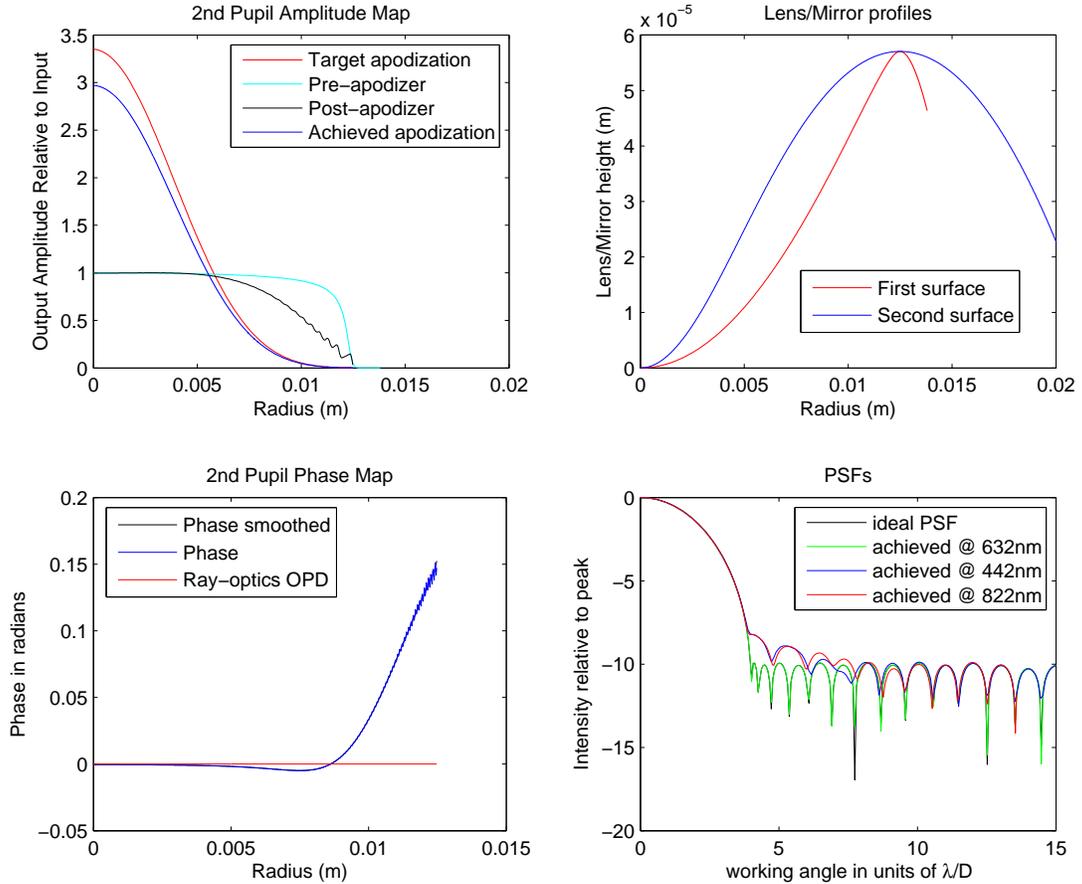}}
\end{center}
\caption{ Analysis of an apodized pupil mapping system using the
S-Huygens approximation with $z = 15D$ and $n=1.5$. {\em Upper-left}
plot shows in red the target high-contrast amplitude profile and in
blue the amplitude profile computed using the S-Huygens
approximation through the apodized pupil mapping system. The other
two curves depict the pre- and post-apodizers. {\em Upper-right}
plot shows the lens profiles, red for the first lens and blue for
the second. The lens profiles $h$ and $\htild$ were computed using a
$5,000$ point discretization. {\em Lower-left} plot shows in red the
computed optical path length $Q_0(\rt)$ and in blue the phase map
computed using the S-Huygens propagation computed using a $5,000$
point discretization. {\em Lower-right} plot shows the PSF computed
at three different wavelengths; the design value, $30\%$ above that
value, and $30\%$ below it. } \label{fig:4}
\end{figure}

Figure \ref{fig:4} shows plots characterizing the performance of an
apodized pupil mapping system analyzed using the techniques
described in this section. The specifications for this system are as
follows. The designed-for wavelength is $632.8$nm. The optical
elements are assumed to be mirrors separated by $0.375$m. The system
is an on-axis system and we therefore make the non-physical
assumption that the mirrors don't obstruct the beam. That is, the
mirrors are invisible except when they are needed. The mirrors take
as input a $0.025$m on-axis beam and produce a $0.025$m
pupil-remapped exit beam. The second mirror is oversized by a factor
of two; that is, its diameter is $0.050$m. The postapodizer ensures
that only the central half contributes to the exit beam. The first
mirror is also oversized appropriately as shown in the upper-right
subplot of Figure \ref{fig:4}. After the second mirror, the exit
beam is brought to a focus. The focal length is $2.5$m. The
lower-right subplot in Figure \ref{fig:4} shows the ideal PSF (in
black) together with the achieved PSF at three wavelengths: at
$70\%$ (green), $100\%$ (blue), and $130\%$ (red) of the design
wavelength. At the design wavelength, the achieved PSF matches the
ideal PSF almost exactly. Note that there is minor degradation at
the other two wavelengths mostly at low spatial frequencies.

We end this section by pointing out that the S-Huygens approximation
given by \eqref{160} is the basis for all subsequent analysis in
this paper. It can be used to compute the propagation between every
pair of consecutive components in apodized pupil mapping and
concentric ring systems. It should be noted that the approximation
does not reduce to the standard Fresnel or Fourier approximations
even when considering such simple scenarios as free-space
propagation of a plane wave or propagation from a pupil plane to an
image plane. Even for these elementary situations, the S-Huygens
approximation is superior to the usual textbook approximations.

\subsection{Propagation of Azimuthal Harmonics}
\label{subsec:Propagation of Azimuthal Harmonics}

In this section, we assume that $E(r,\theta) = E(r) e^{i n \theta}$
for some integer $n$. We refer to such a field as an {\em
$n$th-order azimuthal harmonic}. We will show that an $n$th-order
azimuthal harmonic will remain an $n$th-order azimuthal harmonic
after propagating from the input plane to the output plane described
in the previous section. Only the radial component $E(r)$ changes,
which enables the reduction of the computation from 2D to 1D.
Arbitrary fields can also be propagated, by decomposing them into
azimuthal harmonics and propagating each azimuthal harmonic
separately. Computation is thus greatly simplified even for
arbitrary fields, especially for the case of fields which can be
described by only a few azimuthal harmonics to a high precision,
such as Zernike aberrations, which we consider in subsection
\ref{subsec:Zernike}. This improvement in computation efficiency is
important, because a full 2D diffraction simulation of an apodized
pupil mapping system with the precision of greater than $10^{10}$
typically overwhelms the memory of a mainstream computer. By
reducing the computation from 2D to 1D, however, the entire apodized
pupil mapping system can be simulated with negligible memory
requirements and takes only minutes.

\begin{theorem}
Suppose that the input field in an optical system described by
\eqref{160} is an $n$th-order azimuthal harmonic $\Ein(r,\theta) =
\Ei(r) e^{i n \theta}$ for some integer $n$. Then the output field
is also an $n$th-order azimuthal harmonic $\Eout(\rt,\thetat) =
\Eo(\rt) e^{i n \thetat}$ with radial part given by
\[
    \Eo(\rt)
    =
    \frac{2 \pi i^{n-1} \Aout(\rt)}{\lambda Z}
    \int_0^{\infty} K(r,\rt) \Ei(r)
    J_n\left(\frac{2 \pi r \rt}{\lambda S(\rt,r,0)}\right) \; \Ain(r) r dr .
\]
\end{theorem}
\begin{proof}
We start by substituting the azimuthal harmonic form of $\Ein$
into \eqref{160} and regrouping factors to get
\begin{eqnarray}
  \Eout(\rt,\thetat)
  & = &
    \frac{\Aout(\rt)}{\lambda i Z}
    \int_0^{\infty} K(r,\rt)
    \int_0^{2 \pi}
        e^{2 \pi i \left( -\frac{\rt r \cos(\theta - \thetat)}{S(\rt,r,0)}
                      \right) /\lambda} \Ei(r)e^{i n \theta} d\theta
              \; \Ain(r) r dr \nonumber \\
  & = &
    \frac{\Aout(\rt)}{\lambda i Z}
    e^{i n \thetat}
    \int_0^{\infty} K(r,\rt) \Ei(r)
    \int_0^{2 \pi}
        e^{2 \pi i \left( -\frac{\rt r \cos(\theta - \thetat)}{S(\rt,r,0)}
                      \right) /\lambda}
              e^{i n (\theta-\thetat)}
              d\theta
              \; \Ain(r) r dr . \nonumber \\
\end{eqnarray}
The result then follows from an explicit integration over the $\theta$
variable:
\begin{eqnarray}
  \Eout(\rt,\thetat)
  & = &
    \frac{2 \pi i^{n-1} \Aout(\rt)}{\lambda Z}
    e^{i n \thetat}
    \int_0^{\infty} K(r,\rt) \Ei(r)
    J_n\left(\frac{2 \pi r \rt}{\lambda S(\rt,r,0)}\right)
              \; \Ain(r) r dr \nonumber
\end{eqnarray}
\end{proof}

\subsection{Decomposition of Zernike Aberrations into Azimuthal Harmonics}
\label{subsec:Zernike}
The theorem shows that the full 2D propagation of azimuthal
harmonics can be computed efficiently by evaluating a 1D integral.
However, suppose that the input field is not an azimuthal harmonic,
but something more familiar, such as a $(l,m)$-th Zernike
aberration:

\begin{equation} \label{alpha}
    \Ein(r,\theta) = e^{i\epsilon Z_l^m(r/a) \cos(m \theta)} ,
\end{equation}
where $\epsilon$ is a small number. ($\epsilon/2\pi$ and
$\epsilon/\pi$ are the peak-to-valley phase variations across the
aperture of radius $a$ for $m=0$ and $m\neq0$, respectively.)

%
Recall that the definition of the $n$th-order Bessel function is
\begin{equation} \label{Jn}
    J_n(x) = \frac{1}{2 \pi i^n}
              \int_0^{2 \pi} e^{i x \cos \theta} e^{i n \theta} d \theta .
\end{equation}
From this definition we see that $i^k J_n(x)$ are simply the
Fourier coefficients of $e^{i x \cos(\theta)}$.
Hence, the Fourier series for the complex exponential is given simply by the
so-called {\em Jacobi-Anger expansion}
\begin{equation} \label{jacobi-anger}
    e^{i x \cos \theta} = \sum_{k= -\infty}^{\infty} i^k J_k(x) e^{i k \theta}.
\end{equation}
The Zernike aberration can be decomposed into azimuthal harmonics using
the Jacobi-Anger expansion:
\begin{eqnarray} \label{Zernike}
    e^{i\epsilon Z_l^m(r/a) \cos(m \theta)}
    & = &
    \sum_{k= -\infty}^{\infty} i^k J_k(\epsilon Z_l^m(r/a)) e^{i km
    \theta}\\
    & = &
    J_0(\epsilon Z_l^m(r/a))+\sum_{k= 1}^{\infty} i^k J_k(\epsilon Z_l^m(r/a)) e^{i km \theta}
\end{eqnarray}

Note that
\[
    |J_k(x)| \approx \frac{1}{k!} \left(\frac{x}{2}\right)^k
\]
for $0 \le x \ll 1$.  Hence, if we assume that $\epsilon \sim
10^{-3}$, then the $k$'th term is of the order $10^{-3k}$. The field
amplitude in the high-contrast region of the PSF will be dominated
by the $k=1$ term and be on the order of $10^{-3}$. If we drop terms
of $k=3$ and above, we are introducing an error on the order of
$10^{-9}$ in amplitude. The error in intensity will be dominated by
a cross-product of the $k=3$ and the $k=1$ term, or $10^{-12}$
across the dark region. So, in this case, Zernike aberrations can be
more than adequately modeled using just 3 azimuthal harmonic terms.
For $\epsilon \sim 10^{-2}$, the number of terms goes up to 5 for an
error tolerance of $10^{-12}$. In practice, even this small number
of terms was actually found to be overly conservative.

In order to compute the full 2D response for a given Zernike
aberration, we simply decompose it into a few azimuthal harmonics,
propagate them separately, and sum the results at the end. This
method could also be applied to any arbitrary field.

\section{Simulations} \label{sec:Simulations}

The entire 4-mirror apodized pupil mapping system can be modeled as
the following sequence of 7 steps:

\begin{enumerate}
\item Propagate an input wavefront from the front (flat)
surface of the first pupil mapping lens to the back (flat) surface of the
second pupil mapping lens as described in Section
\ref{subsec:Apodized Pupil Mapping Systems}.
\item Propagate forward a distance $f$.
\item Propagate through a positive lens with focal length $f$ to a focal plane
$f$ units downstream.
\item Multiply by star occulter.
\item Propagate through free-space a distance $f$ then through a positive
lens to recollimate the beam.
\item Propagate forward a distance $f$.
\item Propagate backwards through a pupil mapping system having the same
parameters as the first one.
\end{enumerate}

A similar analysis can be carried out for a concentric ring shaped
pupil system, or even a pure apodization system, as follows:
\begin{enumerate}
\item Choose $\Ain$ to represent either the concentric ring binary mask or
some other azimuthally symmetric apodization.
\item Choose $h$ as appropriate for a focusing lens and let $\htild \equiv 0$.
\item Propagate through this system a distance $f$ to the image plane.
\item Multiply by star occulter.
\end{enumerate}

The theorem can be applied to every propagation step, so that an
azimuthal harmonic will remain an azimuthal harmonic throughout the
entire system. Hence, our computation strategy was to decompose the
input field into azimuthal harmonics, propagate each one separately
through the entire system by repeated applications of the theorem,
and sum them at the very end.

\begin{figure}
\begin{center}
\mbox{\includegraphics[width=5.0in]{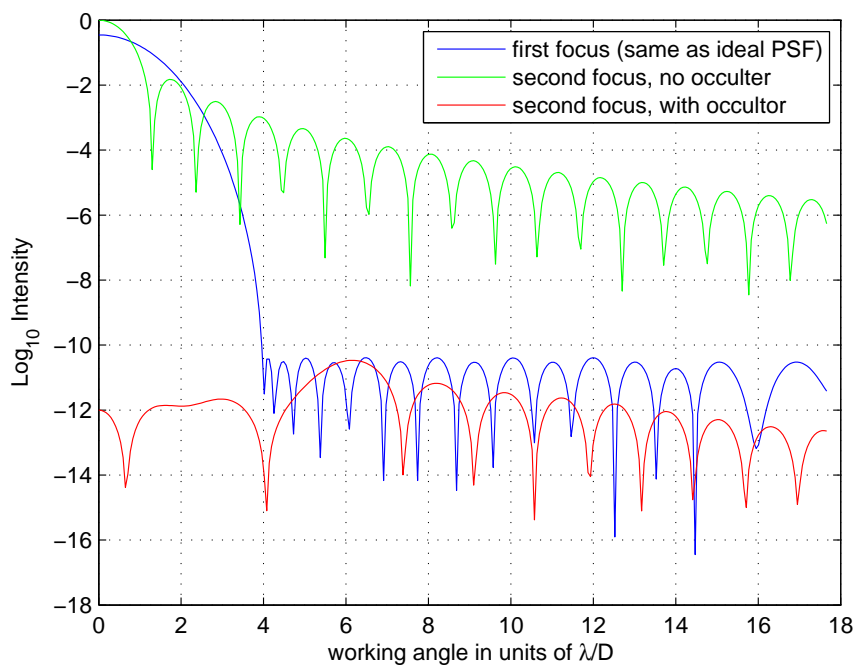}}
\end{center}
\caption{ On-axis PSF at first focus (before occulter) and at second
focus for cases of with and without occulter.  Without the occulter,
the second-focus PSF almost perfectly matches the usual Airy
pattern.  However, with the occulter, the second-focus on-axis PSF
is suppressed by ten orders of magnitude. } \label{fig:5}
\end{figure}

Figure \ref{fig:5} shows a cross section plot of the PSF as it
appears at first focus and second focus in our apodized pupil
mapping system (the first focus plot is indistinguishable from the
case of ideal apodization or concentric ring shaped pupils). There
are two plots for second focus: one with the occulter in place and
one without it. Note that without the occulter, the PSF matches
almost perfectly the usual Airy pattern. With the occulter, the
on-axis light is suppressed by ten orders of magnitude.

\begin{figure}
\begin{center}
\mbox{\includegraphics[width=6.0in]{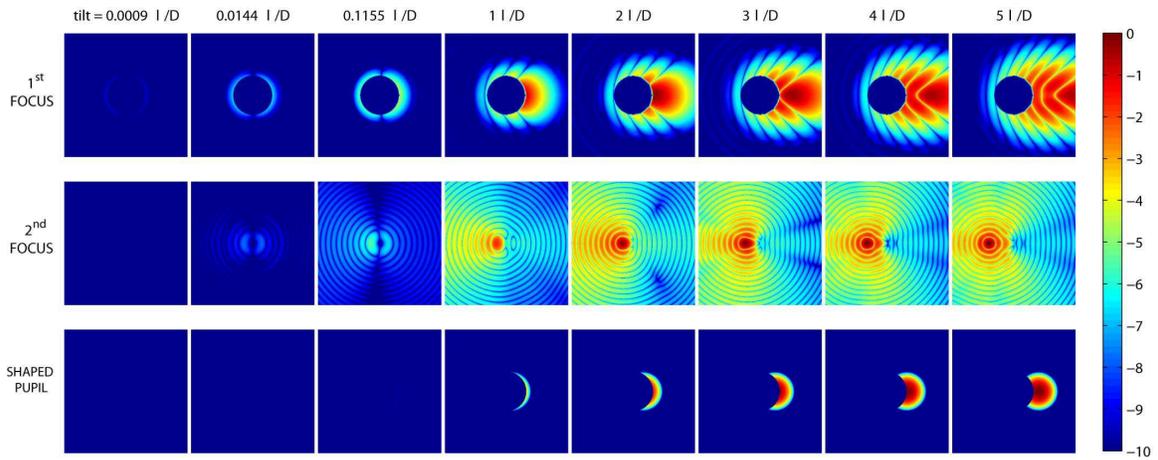}}
\end{center}
\caption{ 2D pictures of planets for apodized pupil mapping
and concentric rings.
{\em First row} shows 2D intensity plots at first focus behind the occulter for
planets at various angles relative to the on-axis star.
Note that the system fails to form a clean image of the planets.
{\em Second row} shows analogous plots at second focus.  Note that the wavefront
for the off-axis planet is mostly restored and the images begin to look like
standard Airy patterns as the angle increases.
{\em Third row} shows analogous plots for a concentric ring mask.
}
\label{fig:6}
\end{figure}

\begin{figure}
\begin{center}
\mbox{\includegraphics[width=6.0in]{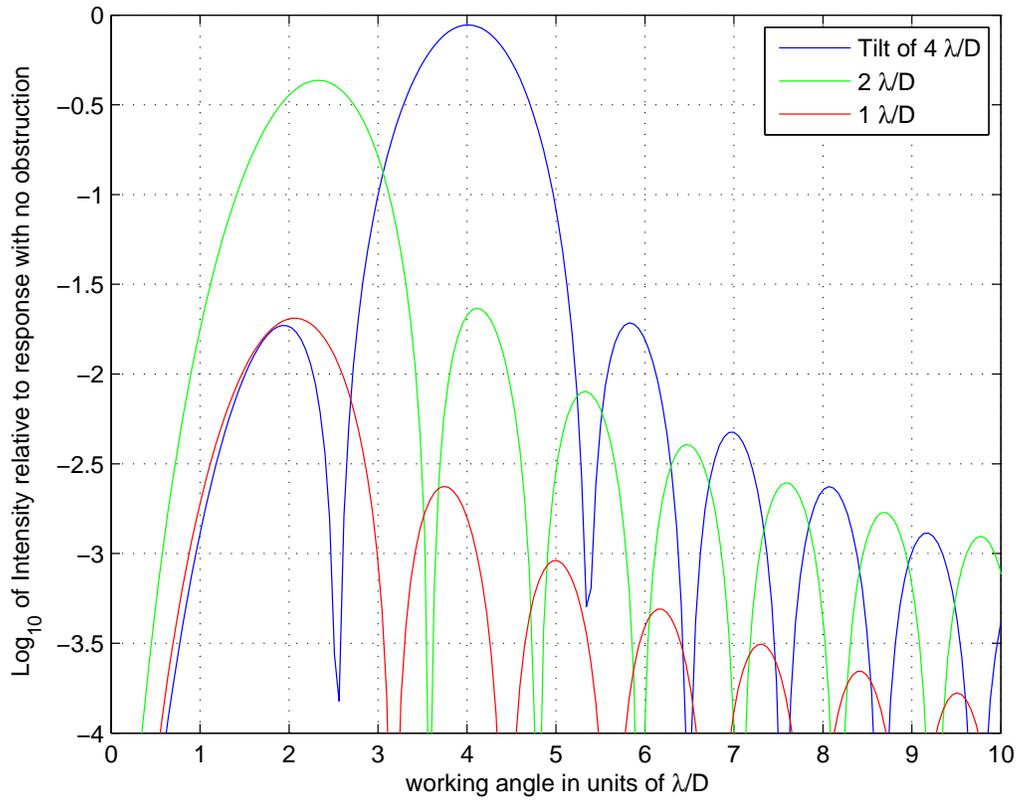}}
\end{center}
\caption{ Cross-sectional plots from the second row plots in Figure
\ref{fig:6}.
Note that for angles of $3 \lambda/D$ and above, the restored PSF looks very
much like an Airy pattern with very little energy attenuation.  However, as
the angle decreases, the pattern begins to distort and the throughput begins
to diminish.
}
\label{fig:7}
\end{figure}

The electric field for a planet is just a slightly tilted and much
fainter field than the field associated with the star.  Hence, the
methods presented here (specifically using the $(1,1)$-Zernike, i.e.
tilt) can be used to generate planet images.
Some such scenarios are shown in Figure \ref{fig:6}. The first row
shows how an off-axis source, i.e. planet, looks at the first focus.
As discussed earlier, at this focal plane off-axis sources do not
form good images.  This is clearly evident in this figure. The
second row shows the planet as it appears at the second image plane,
which is downstream from the reversed pupil mapping system.  In this
case, the off-axis source is mostly restored and the images begin to
look like standard Airy patterns as the angle increases from about
$2 \lambda/D$ outward. Figure \ref{fig:7} shows corresponding cross
sectional plots for the apodized pupil mapping system at the second
focus. The third row in Figure \ref{fig:6} shows how a planet would
appear at a focal plane of a concentric ring shaped pupil system.

\begin{figure}
\begin{center}
\mbox{\includegraphics[width=6.0in]{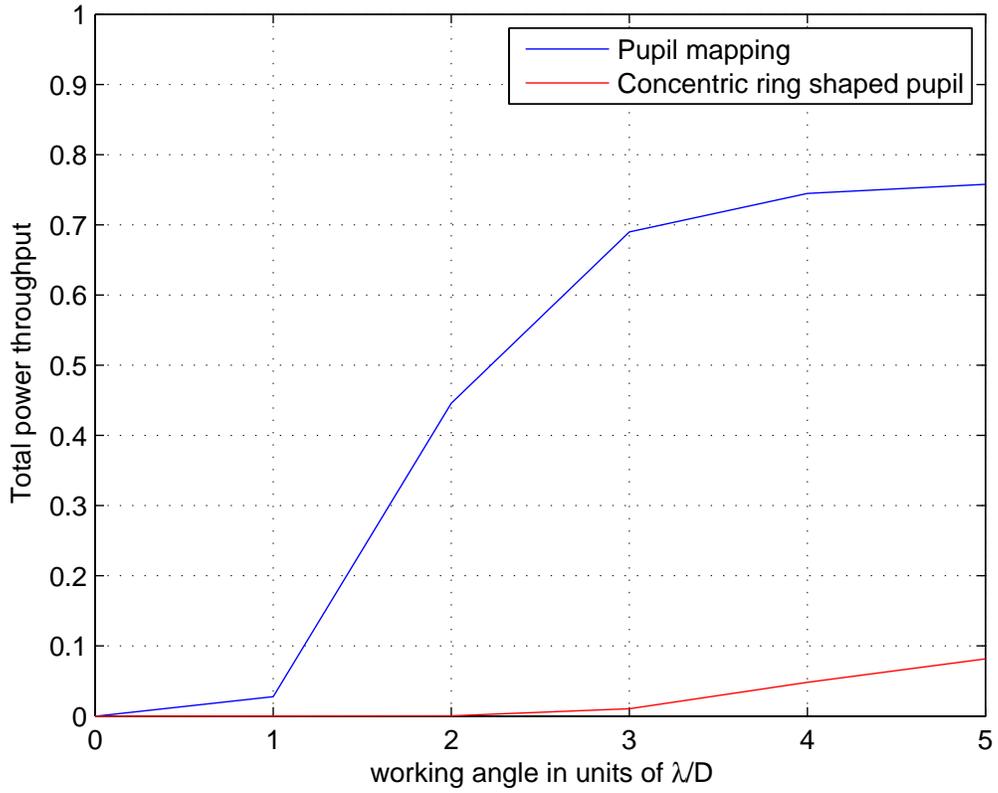}}
\end{center}
\caption{
Off-axis source attenuation as a function of angle.
Note that the $50\%$ point occurs at about $2.5 \lambda/D$.
}
\label{fig:8}
\end{figure}

Figure \ref{fig:8} shows how the off-axis source is attenuated as a
function of the angle from optical axis, for the case of our
apodized pupil mapping system (at second focus) and the concentric
ring coronagraph. For the case of apodized pupil mapping, the $50\%$
point occurs at about $2.5 \lambda/D$.

\begin{figure}
\begin{center}
\mbox{\includegraphics[width=6.0in]{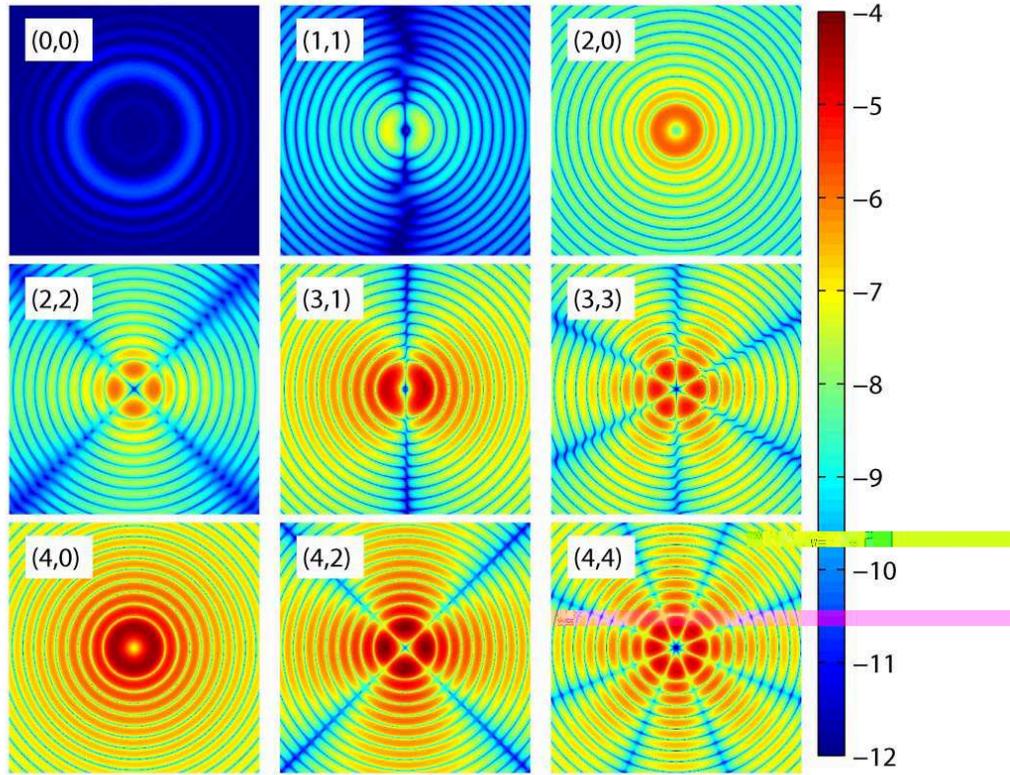}}
\end{center}
\caption{
Apodized pupil mapping sensitivities to the first nine Zernike aberrations.
In each case, the rms error is $1/100$th wave.
The plots correspond to: Piston $(0,0)$, Tilt $(1,1)$, Defocus
$(2,0)$, Astigmatism $(2,2)$ Coma $(3,1)$, Trefoil $(3,3)$,
Spherical Aberration $(4,0)$, Astigmatism 2nd Order $(4,2)$,
Tetrafoil $(4,4)$.
}
\label{fig:9}
\end{figure}

\begin{figure}
\begin{center}
\mbox{\includegraphics[width=6.0in]{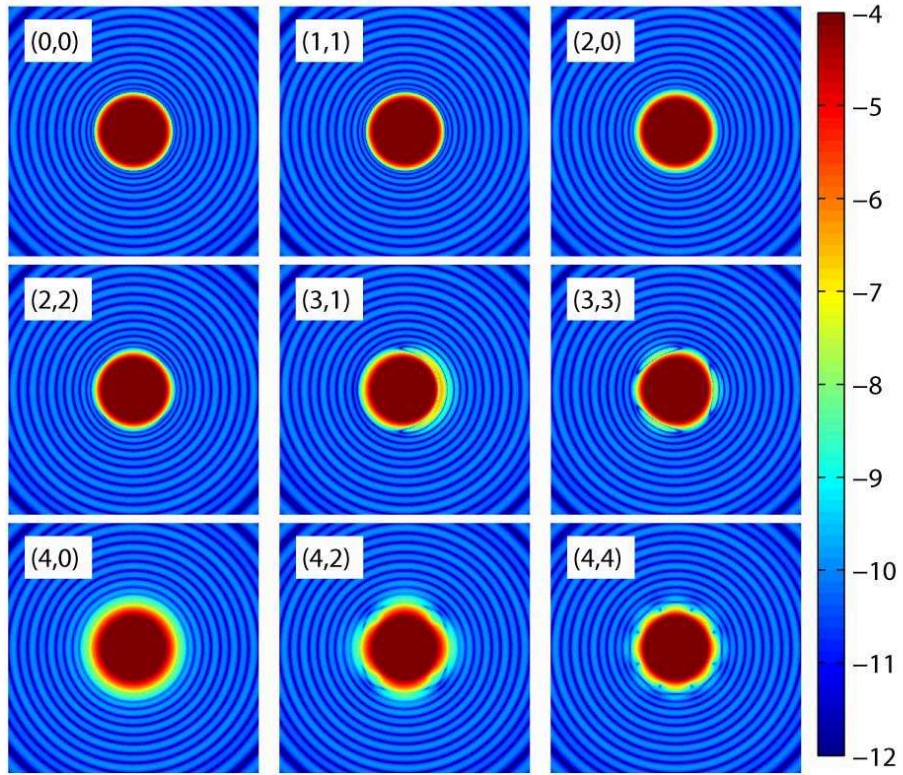}}
\end{center}
\caption{
Concentric-ring mask sensitivities to the first nine Zernike aberrations.
}
\label{fig:10}
\end{figure}

Figures \ref{fig:9} and \ref{fig:10} show the distortions/leakage
from an on-axis source in the presence of various Zernike
aberrations, for apodized pupil mapping and the concentric ring
shaped pupil systems, respectively. The Zernike aberrations are
assumed to be $1/100$th wave rms.

Figure \ref{fig:11} shows the corresponding cross-section
sensitivity plots for both the apodized pupil mapping system and the
concentric ring shaped pupil system. From this plot it is easy to
see both the tighter inner working angle of apodized pupil mapping
systems as well as their increased sensitivity to wavefront errors.
Finally, Figure \ref{fig:12} demonstrates contrast degradation
measured at three angles, $2$, $4$, and $8 \lambda/D$, as a function
of severity of the Zernike wavefront error.  The rms error is
expressed in waves.

\section{Conclusions} \label{sec:Conclusions}

\begin{figure}
\begin{center}
\mbox{\includegraphics[width=6.0in]{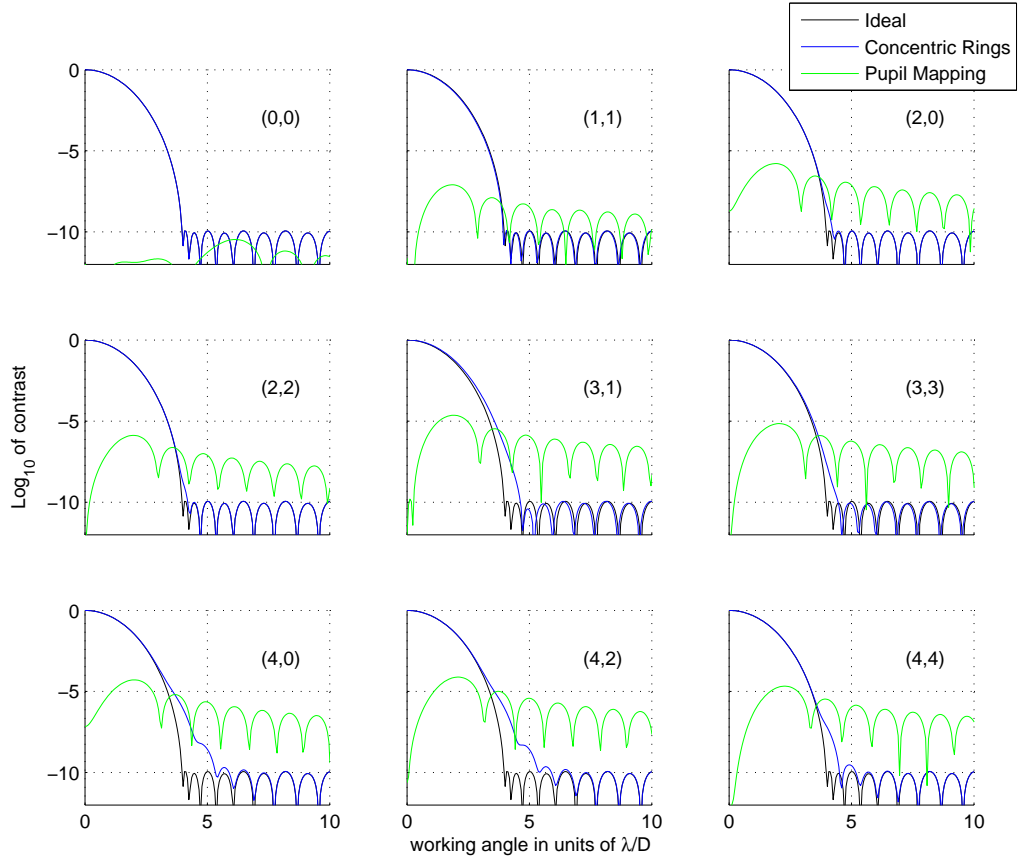}}
\end{center}
\caption{ Radial profiles associated with the previous two Figures
and overlayed one on the other.  The green plots are for apodized
pupil mapping whereas the blue plots are for the concentric ring
mask. } \label{fig:11}
\end{figure}

\begin{figure}
\begin{center}
\mbox{\includegraphics[width=6.0in]{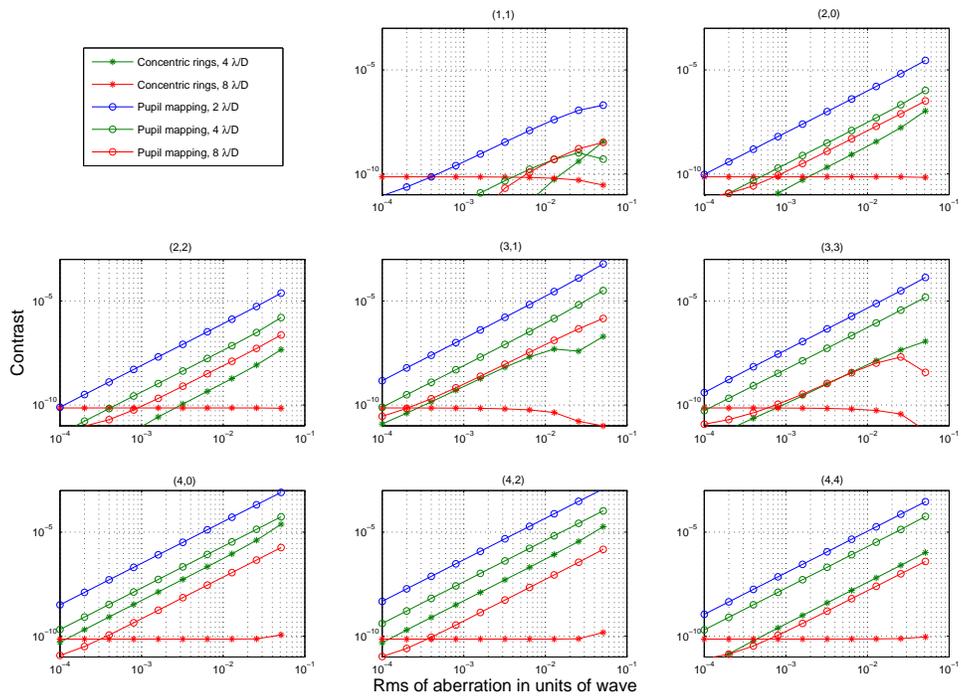}}
\end{center}
\caption{
Contrast degradation measured at three angles, $2$, $4$, and $8 \lambda/D$
as a function of severity of the Zernike wavefront error measured in waves.
}
\label{fig:12}
\end{figure}

We have presented an efficient method for calculating the
diffraction of aberrations through optical systems such as apodized
pupil mapping and shaped pupil coronagraphs. We presented an example
for both systems and computed their off-axis responses and
aberration sensitivities. Figures \ref{fig:11} and \ref{fig:12} show
that our particular apodized pupil mapping system is more sensitive
to low order aberrations than the concentric ring masks. That is,
contrast and IWA degrade more rapidly with increasing rms level of
the aberrations. Thus, for a particular telescope, our pupil mapping
system will achieve better throughput and inner working angle, but
suffer greater aberration sensitivity.

We note that there is a spectrum of apodized pupil mapping systems,
out of which we selected but one example. The two extremes, pure
apodization and pure pupil mapping, both have serious drawbacks.  On
the one end, pure apodization loses almost an order of magnitude in
throughput and suffers from an unpleasantly large IWA.  At the other
extreme, pure pupil mapping fails to achieve the required high
contrast due to diffraction effects. There are several points along
this spectrum that are superior to the end points.  We have focused
on just one such point, which is similar to the design suggested by
\cite{GPGMRW05}. We leave it to future work to determine if this is
the best design point. For example, clearly one can improve the
aberration sensitivity by relaxing the inner working angle and
throughput requirements. Such analysis is beyond the scope of this
paper, but we have provided here the tools to analyze the
sensitivity of these kinds of designs.

{\bf Acknowledgements.}
This research was partially performed for the
Jet Propulsion Laboratory, California Institute of Technology,
sponsored by the National Aeronautics and Space Administration as part of
the TPF architecture studies and also under JPL subcontract number 1260535.
The third author also received support from the ONR (N00014-05-1-0206).

\bibliography{../lib/refs}   
\bibliographystyle{plainnat}   


\end{document}